\documentclass[twoside]{book}
\usepackage{amsmath}
\usepackage{amsfonts}
\usepackage{amssymb}
\usepackage{graphicx}		

 \makeatletter
\ifx\UNDEF\mail\def\mail{ }\else\fi
\ifx\UNDEF\prange\def\prange{0 0}\else\fi

\gdef\@empty{}
\def\Mail#1 #2 {\gdef\thecontact{#1}\gdef\theaddr{#2}}
\def\Range#1 #2 {\gdef\thefirstpage{#1}\gdef\thelastpage{#2}}
{\let\'\mail \expandafter\Mail\' }  
{\let\'\prange \expandafter\Range\' }   
 \gdef\@shtitle{\relax}
 \long\def\shtitle#1{\gdef\@shtitle{#1}}
 \long\def\author#1{\gdef\@author{#1}}
 \def\affil#1{\par\noindent{\rm#1\par}}
 \gdef\@abstract{}
 \long\def\abstract#1{\gdef\@abstract{#1}}

 \def\maketitle{\thispagestyle{empty}\chapter{\@title}}
 \renewcommand\chapter{\if@openright\cleardoublepage\else\clearpage\fi
                    \thispagestyle{empty}%
                    \global\@topnum\z@
                    \@afterindentfalse
                    \secdef\@chapter\@schapter}
 \def\@makechapterhead#1{%
  \vspace*{50\p@}%
  {\parindent \z@ \raggedleft \normalfont
    \ifnum \c@secnumdepth >\m@ne
      \if@mainmatter
        \par\nobreak
        \vskip 20\p@
      \fi
    \fi
    \interlinepenalty\@M
    \Huge \bfseries #1\par\nobreak
    \vskip.25in
    \large\bfseries\@author\par\nobreak
    \vskip 40\p@}
    \ifx\@abstract\@empty\else{\small\@abstract\par\vskip20\p@}\fi
  }

\DeclareRobustCommand\em
        {\@nomath\em \ifdim \fontdimen\@ne\font >\z@
                       \upshape \else \slshape \fi}

\def\@begintheorem#1#2{\sl \trivlist \item[\hskip \labelsep{\bf #1\ #2}]}
\def\@opargbegintheorem#1#2#3{\sl \trivlist
     \item[\hskip \labelsep{\bf #1\ #2\ (#3)}]}

  \def\@arabic#1{\number #1} 

\long\def\@makecaption#1#2{
    \vskip\abovecaptionskip
    \sbox\@tempboxa{{\small {\bf #1}: #2}}%
    \ifdim\wd\@tempboxa>\hsize
        {\small {\bf #1}: #2\par}
    \else
       \global\@minipagefalse
       \hbox to\hsize{\hfil\box\@tempboxa\hfil}
    \fi
    \vskip \belowcaptionskip}

\def\figstrut#1{\hbox to\linewidth{\vrule height#1\hfill}}

\renewenvironment{thebibliography}[1]
     {\section*{\bibname
        \@mkboth{\MakeUppercase\bibname}{\MakeUppercase\bibname}}%
      \list{\@biblabel{\@arabic\c@enumiv}}%
           {\settowidth\labelwidth{\@biblabel{#1}}%
            \leftmargin\labelwidth
            \advance\leftmargin\labelsep
            \@openbib@code
            \usecounter{enumiv}%
            \let\p@enumiv\@empty
            \renewcommand\theenumiv{\@arabic\c@enumiv}}%
      \sloppy
      \clubpenalty4000
      \@clubpenalty \clubpenalty
      \widowpenalty4000%
      \sfcode`\.\@m}
     {\def\@noitemerr
       {\@latex@warning{Empty `thebibliography' environment}}%
      \endlist}
\makeatother

 \shtitle{Inferring Diversity}	
 \title{Inferring Diversity: Life After Shannon}
 \author{Adom Giffin\affil{Department of Physics\\ University at Albany--SUNY\\ Albany, NY 12222,USA\\ physics101@gmail.com}}
 \abstract{The diversity of a community that cannot be fully counted must be inferred.
The two preeminent inference methods are the MaxEnt method, which uses
information in the form of constraints and Bayes' rule which uses
information in the form of data. It has been shown that these two methods
are special cases of the method of Maximum (relative) Entropy (ME). We
demonstrate how this method can be used as a measure of diversity that not
only reproduces the features of Shannon's index but exceeds them by allowing
more types of information to be included in the inference. A specific
example is solved in detail. Additionally, the entropy that is found is the same form as the thermodynamic entropy.}

\begin{document}           
\maketitle

\section{Introduction}

\label{intro}

Diversity is a concept that is used in many fields to describe the
variability of different entities in a group. In ecology, the Shannon
entropy \cite{Shannon48} and Simpson's index \cite{Simpson49} are the
predominate measures of diversity. In this paper we focus on the Shannon
entropy for two reasons: First, it has been shown that Simpson's index is an
approximation of Shannon's \cite{Guiasu02}. Second, Shannon's entropy is
closely tied to many other areas of research, such as information theory and
physics.

It is often the case that the species in a community cannot be fully
counted. In this case, when one has incomplete information, one must rely on
methods of inference. The two preeminent inference methods are the MaxEnt 
\cite{Jaynes57} method, which has evolved to a more general method, the
method of Maximum (relative) Entropy (ME) \cite%
{ShoreJohnson80,Skilling88,CatichaGiffin06} and Bayes' rule. The choice
between the two methods has traditionally been dictated by the nature of the
information being processed (either constraints or observed data). However,
it has been shown that one can accommodate both types of information in one
method, ME \cite{GiffinCaticha07}. The purpose of this paper is to
demonstrate how the ME method can be used as a measure of diversity that is
able to include more information that Shannon's measure allows.

Traditionally when confronted with a community whose count is incomplete,
the frequency of the species that are counted are used to calculate the
diversity. The frequency is used because it represents an estimate of the
probability of finding a particular species in the community. However, the
frequency is not equivalent to the probability \cite{Sivia96} and as such is
a poor estimate. Fortunately, there are much better methods for estimating
or inferring the probability such as MaxEnt and Bayes. Even more fortunate
is that the new ME method can reproduce every aspect of Bayesian and MaxEnt
inference \emph{and} tackle problems that the two methods alone could not
address.

We start by showing a general example of the ME method by inferring a
probability with two different forms of information: expected values%
\footnote{%
For simplicity we will refer to these expected values as \emph{moments}
although they can be considerably more general.} and data, \emph{%
simultaneously}. The solution resembles Bayes' Rule. In fact, if there are
no moment constraints then the method produces Bayes rule \emph{exactly}. If
there is no data, then the MaxEnt solution is produced.

Finally we solve a toy ecological problem and discuss the diversity
calculated by using Shannon's entropy and the diversity calculated by the ME
method. This illustrates the many advantages to using the ME method.

\section{Simultaneous updating}

\label{sec:2}

Our first concern when using the ME method to update from a prior to a
posterior distribution\footnote{%
In Bayesian inference, it is assumed that one always has a prior probability
based on some prior information. When new information is attained, the old
probility (the prior) is \emph{updated} to a new probability (the
posterior). If one has no prior information, then one uses an \emph{ignorant}
prior \cite{Gelman04}.} is to define the space in which the search for the
posterior will be conducted. We wish to infer something about the values of
one or several quantities, $\theta \in \Theta $, on the basis of three
pieces of information: prior information about $\theta $ (the prior), the
known relationship between $x$ \emph{and} $\theta $ (the model), and the
observed values of the data $x\in \mathcal{X}$. Since we are concerned with
both $x$ \emph{and} $\theta $, the relevant space is neither $\mathcal{X}$
nor $\Theta $ but the product $\mathcal{X}\times \Theta $ and our attention
must be focused on the joint distribution $P(x,\theta )$. The selected joint
posterior $P_{\text{new}}(x,\theta )$ is that which maximizes the entropy,%
\begin{equation}
S[P,P_{\text{old}}]=-\int dxd\theta ~P(x,\theta )\log \frac{P(x,\theta )}{P_{%
\text{old}}(x,\theta )}~,  \label{entropy}
\end{equation}%
subject to the appropriate constraints. $P_{\text{old}}(x,\theta )$ contains
our prior information which we call the \emph{joint prior}. To be explicit,%
\begin{equation}
P_{\text{old}}(x,\theta )=P_{\text{old}}(\theta )P_{\text{old}}(x|\theta )~,
\label{joint prior}
\end{equation}%
where $P_{\text{old}}(\theta )$ is the traditional Bayesian prior and $P_{%
\text{old}}(x|\theta )$ is the likelihood. It is important to note that they 
\emph{both} contain prior information. The Bayesian prior is defined as
containing prior information. However, the likelihood is not traditionally
thought of in terms of prior information. Of course it is reasonable to see
it as such because the likelihood represents the model (the relationship
between $\theta $ and $x)$ that has already been established. Thus we
consider both pieces, the Bayesian prior and the likelihood to be \emph{prior%
} information.

The new information is the \emph{observed data}, $x^{\prime }$, which in the
ME framework must be expressed in the form of a constraint on the allowed
posteriors. The family of posteriors that reflects the fact that $x$ is now
known to be $x^{\prime }$ is such that%
\begin{equation}
C_{1}:P(x)=\int d\theta ~P(x,\theta )=\delta (x-x^{\prime })~.  \label{data}
\end{equation}%
This amounts to an \emph{infinite} number of constraints: there is one
constraint on $P(x,\theta )$ for each value of the variable $x$ and each
constraint will require its own Lagrange multiplier $\lambda (x)$.
Furthermore, we impose the usual normalization constraint, 
\begin{equation}
\int dxd\theta ~P(x,\theta )=1~,  \label{Normalization}
\end{equation}%
and include additional information about $\theta $ in the form of a
constraint on the expected value of some function $f(\theta )$\footnote{%
Including an additional constraint in the form of $\int dxd\theta P(x,\theta
)g(x)=\left\langle g\right\rangle =G$ could only be used when it does not
contradict the data constraint (\ref{data}). Therefore, it is redundant and
the constraint would simply get absorbed when solving for $\lambda (x)$.}, 
\begin{equation}
C_{2}:\int dxd\theta \,P(x,\theta )f(\theta )=\left\langle f(\theta
)\right\rangle =F~.  \label{moment}
\end{equation}%
We emphasize that constraints imposed at the level of the prior need not be
satisfied by the posterior. What we do here differs from the standard
Bayesian practice in that we \emph{require} the constraint to be satisfied
by the posterior distribution.

Maximize (\ref{entropy}) subject to the above constraints, 
\begin{equation}
\delta \left\{ 
\begin{array}{c}
S+\alpha \left[ \int dxd\theta P(x,\theta )-1\right] \\ 
+\beta \left[ \int dxd\theta P(x,\theta )f(\theta )-F\right] \\ 
+\int dx\lambda (x)\left[ \int d\theta P(x,\theta )-\delta (x-x%
{\acute{}}%
)\right]%
\end{array}%
\right\} =0~,  \label{max}
\end{equation}%
yields the joint posterior,%
\begin{equation}
P_{\text{new}}(x,\theta )=P_{\text{old}}(x,\theta )\frac{e^{\lambda
(x)+\beta f(\theta )}}{Z}~,
\end{equation}%
where $Z$ is determined by using (\ref{Normalization}),%
\begin{equation}
Z=e^{-\alpha +1}=\int dxd\theta e^{\lambda (x)+\beta f(\theta )}P_{\text{old}%
}(x,\theta )
\end{equation}%
and the Lagrange multipliers $\lambda (x)$ are determined by using (\ref%
{data})%
\begin{equation}
e^{\lambda (x)}=\frac{Z}{\int d\theta e^{\beta f(\theta )}P_{\text{old}%
}(x,\theta )}\delta (x-x%
{\acute{}}%
)~.
\end{equation}%
The posterior now becomes%
\begin{equation}
P_{\text{new}}(x,\theta )=P_{\text{old}}(x,\theta )\delta (x-x%
{\acute{}}%
)\frac{e^{\beta f(\theta )}}{\zeta (x,\beta )}~,  \label{Posterior-Both}
\end{equation}%
where $\zeta (x,\beta )=\int d\theta e^{\beta f(\theta )}P_{\text{old}%
}(x,\theta ).$

The Lagrange multiplier $\beta $ is determined by first substituting the
posterior into (\ref{moment}),%
\begin{equation}
\int dxd\theta \left[ P_{\text{old}}(x,\theta )\delta (x-x%
{\acute{}}%
)\frac{e^{\beta f(\theta )}}{\zeta (x,\beta )}\right] f(\theta )=F~.
\end{equation}%
Integrating over $x$ yields,%
\begin{equation}
\frac{\int d\theta e^{\beta f(\theta )}P_{\text{old}}(x^{\prime },\theta
)f(\theta )}{\zeta (x^{\prime },\beta )}=F~,
\end{equation}%
where $\zeta (x,\beta )\rightarrow \zeta (x^{\prime },\beta )=\int d\theta
e^{\beta f(\theta )}P_{\text{old}}(x^{\prime },\theta )$. Now $\beta $ can
be determined by%
\begin{equation}
\frac{\partial \ln \zeta (x^{\prime },\beta )}{\partial \beta }=F~.
\label{F}
\end{equation}

The final step is to marginalize the posterior, $P_{\text{new}}(x,\theta )$
over $x$ to get our updated probability,%
\begin{equation}
P_{\text{new}}(\theta )=P_{\text{old}}(x^{\prime },\theta )\frac{e^{\beta
f(\theta )}}{\zeta (x^{\prime },\beta )}
\end{equation}%
Additionally, this result can be rewritten using the product rule as%
\begin{equation}
P_{\text{new}}(\theta )=P_{\text{old}}(\theta )P_{\text{old}}(x^{\prime
}|\theta )\frac{e^{\beta f(\theta )}}{\zeta ^{\prime }(x^{\prime },\beta )}~,
\end{equation}%
where $\zeta ^{\prime }(x^{\prime },\beta )=\int d\theta e^{\beta f(\theta
)}P_{\text{old}}(\theta )P_{\text{old}}(x^{\prime }|\theta ).$ The right
side resembles Bayes theorem, where the term $P_{\text{old}}(x^{\prime
}|\theta )$ is the standard Bayesian likelihood and $P_{\text{old}}(\theta )$
is the prior. The exponential term is a \emph{modification} to these two
terms. Notice when $\beta =0$ (no moment constraint) we recover Bayes' rule.
For $\beta \neq 0$ Bayes' rule is modified by a \textquotedblleft
canonical\textquotedblright\ exponential factor.

It must be noted that MaxEnt has been traditionally used for obtaining a
prior for use in Bayesian statistics. When this is the case, the updating is
sequential. This is not the case here where both types of information are
processed simultaneously. In the sequential updating case, the multiplier $%
\beta $ is chosen so that the posterior $P_{\text{new}}$ only satisfies $%
C_{2}$. In the simultaneous updating case the multiplier $\beta $ is chosen
so that the posterior $P_{\text{new}}$ satisfies both $C_{1}$ and $C_{2}$ or 
$C_{1}\wedge C_{2}$ \cite{GiffinCaticha07}.

\section{Inference in Ecology}

\label{Example1}

In the following sections we will discuss the traditional way diversity is
measured and the way it is measured using ME. This will be done by examining
a simple example and comparing the two methods. In addition, we will show
how the ME method could include information that the traditional method
cannot.

The general information for the example is as follows: There are $k$ types
of plants in a forest. A portion of the forest is examined and the amount of
each species is counted where $m_{1},m_{2}\ldots m_{k}$ represents the
counts of each species and $n$ represents the total count so that $%
n=\sum\nolimits_{i}^{k}m_{i}.$ Additionally, we know from biological
examination that one species, $s_{2}$ and another species, $s_{5}$ are
codependent. Perhaps they need each others pollen in such supply that they
cannot exist unless there are on the average, twice the number of $s_{2}$ as
compared to $s_{5}.$

\subsection{Traditional Diversity}

\label{ExampleT}

We calculate the Shannon diversity by using Shannon's entropy as follows,%
\begin{equation}
S_{Tradtional}=-\sum_{i}^{k}p_{i}\log p_{i~,}  \label{entropyT}
\end{equation}%
where $p_{i}=m_{i}/n.$ The problem with using this method is not in the
method itself but with the reason it is being used. If the purpose of using
this method was to measure the diversity of the portion that was counted
then the method is acceptable. However, if the purpose of the method is to
estimate or infer the diversity of the \emph{whole} forest, then it is a
poor estimate. First, $p_{i}$ is meant to represent the probability of
finding the $ith$ species in the forest. As previously stated, the frequency
of the sample is not equivalent to the probability. In fact, it is the
expected value of the frequency that is equivalent to the probability, $%
\left\langle F\right\rangle =p$ \cite{Sivia96}. It would only make sense to
use the frequency as an estimate of the probability when $n$ is very large
(i.e. $n\rightarrow \infty $) but this is not usually the case. Second, the
diversity of two samples that have the same ratio of frequencies will be the
same. Therefore this measure does not reflect the abundance of the species.
This might be a desirable feature \cite{Guiasu02}. Third, there is no clear
way to process the information about the codependence using Shannon's
entropy.

\subsection{ME Diversity}

\label{ExampleME}

Here we intend to use a better method to estimate or infer $p_{i}$ and that
method is the ME method. The first task is to realize that the correct
mathematical model for the probability of getting a particular species where
the information that we have is the number of species counted is a
multinomial distribution. The probability of finding $k$ species in $n$
counts which yields $m_{i}$ instances for the $i^{th}$ species is%
\begin{equation}
P_{\text{old}}(m|p,n)=P_{\text{old}}(m_{1}\ldots m_{k}|p_{1}\ldots p_{k},n)=%
\frac{n!}{m_{1}!\ldots m_{k}!}p_{1}^{m_{1}}\ldots p_{k}^{m_{k}}~,
\label{multinomial}
\end{equation}%
where $m=(m_{1},\ldots ,m_{k})$ with $\sum\nolimits_{i=1}^{k}m_{i}=n$, and $%
p=(p_{1},\ldots ,p_{k})$ with $\sum\nolimits_{i=1}^{k}p_{i}=1$. The general
problem is to infer the parameters $p$ on the basis of information about the
data, $m^{\prime }.$ Here we see the first advantage with using the ME
diversity; we allow for fluctuations in our inference by looking at a
distribution of $p^{\prime }s$ as opposed to claiming that we know the
"true" $p.$

Additionally we can include information about the codependence by using the
following general constraint,%
\begin{equation}
\left\langle f(p)\right\rangle =F\quad \text{where}\quad
f(p)=\sum\nolimits_{i}^{k}f_{i}p_{i}~,  \label{cP1}
\end{equation}%
where $f_{i}$ is used to represent the codependence. For our example, on the
average, we will find twice the number of $s_{2}$ as compared to $s_{5}$
thus, \emph{on the average}, the probability of finding one of the species
will be twice that of the other, $\left\langle p_{2}\right\rangle
=2\left\langle p_{5}\right\rangle $. In this case, $f_{2}=1,$ $f_{5}=-2$ and 
$f_{i\neq \left( 2,5\right) }=F=0.$

Next we need to write the data (counts) as a constraint which in general is%
\begin{equation}
P(m|n)=\delta _{mm^{\prime }}~,  \label{cP2}
\end{equation}%
where $m^{\prime }=\{m_{1}^{\prime },\ldots ,m_{k}^{\prime }\}.$ Finally we
write the appropriate entropy to use,%
\begin{equation}
S[P,P_{\text{old}}]~\text{=}-\sum\limits_{m}\int dpP(m,p|n)\log \frac{%
P(m,p|n)}{P_{\text{old}}(m,p|n)}~,  \label{entropyP}
\end{equation}%
where%
\begin{equation}
\sum\limits_{m}=\sum\limits_{m_{1}\ldots m_{k}=0}^{n}\delta
(\sum\nolimits_{i=1}^{k}m_{i}-n)~,
\end{equation}%
and%
\begin{equation}
\int dp=\int dp_{1}\ldots dp_{k}\,\delta \left(
\sum\nolimits_{i=1}^{k}p_{i}-1\right) ~,
\end{equation}%
and where $P_{\text{old}}(m,p|n)=P_{\text{old}}(p|n)P_{\text{old}}(m|p,n).$
The prior $P_{\text{old}}(p)$ is not important for our current purpose so
for the sake of definiteness we can choose it flat for our example (there
are most likely better choices for priors). We then maximize this entropy
with respect to $P(m,p|n)$ subject to normalization and our constraints
which after marginalizing over $m^{\prime }$ yields,%
\begin{equation}
P(p)=P_{\text{old}}(m^{\prime }|p,n)\frac{e^{\beta f(p)}}{\zeta }~,
\label{posterior P}
\end{equation}%
where%
\begin{equation}
\zeta =\int dp\,e^{\beta f(p)}P_{\text{old}}(m^{\prime }|p,n)\quad \text{and}%
\quad F=\frac{\partial \log \zeta }{\partial \beta }~.  \label{zeta P}
\end{equation}

The probability distribution $P(p)$ has sometimes been criticized for being
too strange. The idea of getting a probability of a probability may seem
strange at first but makes absolute sense. We do not know the "true"
distribution of species, $p_{i}.$ Therefore it seems natural to express our
knowledge with some uncertainty in the form of a distribution. Notice that
if one has no information relating the species then $\beta =0.$

Finally by substituting (\ref{posterior P}) into (\ref{entropyP}), and using
our constraints (\ref{cP1}) and (\ref{cP2}) we introduce our new general
measure for diversity, 
\begin{equation}
S_{ME}=\log \zeta -\beta F~.  \label{entropyME}
\end{equation}

\section{Conclusions}

\label{Conclusions}

Diversity is an important concept in many fields. In this paper we provided
a toy example of how ME would be used as a measure of diversity that may
simulate real world situations. By using the multinomial, we not only
properly infer $p$ so that fluctuations are represented, we get the
additional bonus of having the abundance of the species represented in the
measure. It is critical to note that our diversity, $S_{ME}$ satisfies all
of Pielou's axioms \cite{Pielou75}.

This of course could all be done with only using Bayes to infer $p.$
However, by using the ME method we can include additional information
allowing to go beyond what Bayes' rule and MaxEnt methods alone could do.
Therefore, we would like to emphasize that anything one can do with Bayesian
or MaxEnt methods, one can now do with ME. Additionally, in ME one now has
the ability to apply additional information that Bayesian or MaxEnt methods
could not. Further, any work done with Bayesian techniques can be
implemented into the ME method directly through the joint prior.

Although Shannon had discovered the entropy that bears his name quite
independently of thermodynamic considerations, it nevertheless is directly
proportional to the thermodynamic entropy. The realization that the ME
diversity is of the exact same form as the thermodynamic entropy\footnote{%
The thermodynamic entropy is actually, $S=$ $\log \zeta $ $+\beta F$. The
fact that our entropy (\ref{entropyME}) has a $-\beta F$ is a reflection of
our choice to add our Lagrange multipliers in (\ref{max}) as opposed to
subtracting them as is the case in thermodynamics. However, this is trivial
because when one solves for $\beta $ in (\ref{F}) the sign will be accounted
for. Thus, if the Lagrange multiplier was subtracted, the solution to (\ref%
{F}) would be $-F$ and the entropy would have a $+\beta F.$} is of no small
consequence. All of the concepts that thermodynamics utilizes can now also
be utilized in ecology, whether it be energy considerations or equilibrium
conditions, etc.

To see a detailed method for calculating $\zeta ,$ see \cite{GiffinCaticha07}%
, for a numeric example, see \cite{GiffinEcon07} and for an example of what
do when one knows that there are species in the forest but simply have not
been counted (perhaps they are rare), see \cite{GiffinAgent07}.

\bigskip

\noindent \textbf{Acknowledgements:} We would like to acknowledge many
valuable discussions with A. Caticha. Presented at the 7th International Conference on Complex Systems, Boston, 2007.

\end{document}